\documentclass[twoside]{LCWS11}
\usepackage[latin1]{inputenc}
\usepackage[dvips]{graphicx,epsfig,color}
\usepackage{wrapfig,rotating}
\usepackage{amssymb,amsmath,array}
\usepackage{slashbox}
\begin{document}
\title{Developement of readout ASIC for FPCCD vertex detector} 
\author{Eriko Kato$^1$ Hisao Sato$^2$ Hirokazu Ikeda$^3$  Yasuhiro Sugimoto$^4$\\
 Kennosuke Itagaki$^1$,  Tomoyuki Saito$^1$,Akiya Miyamoto$^4$, Yosuke Takubo$^4$ and Hitoshi Yamamoto$^1$
\vspace{.3cm}\\
1- Department of Physics, Tohoku University, Sendai 980-8578- Japan
\vspace{.1cm}\\
2- Department of Physics, Shinshu University, Matsumoto- 390-8621, Japan\\
3- JAXA, Japan Aerospace Exploration Agency Organization, Sagamihara, Kanagawa 252-5210, Japan\\
4- High Energy Accelerator Research Organization (KEK),Tsukuba 305-0801, Japan
}

\maketitle

\begin{abstract}

\indent One of candidates for the International Linear Collider(ILC)'s vertex detector is the Fine Pixel CCD (FPCCD) with a pixel size of $5 \times 5~ (\mu m ^{2} )$. Sensor and readout systems are currently being studied and prototypes have been developed. In this paper we will report on the performance of latest developed readout ASIC prototype as well as the outline of the design strategy for the next ASIC  prototype.
\end{abstract}

\section{Introduction}

\indent One of the most important physics that the ILC is aiming to study is the verification of the Higgs mechanism, which is reflected in the Higgs boson couplings. The measurement of these various couplings requires accurate particle identification, the most crucial one being the separation between b-quarks and  c-quarks.
In order to realize ILC's highly efficient flavor tagging, we need a vertex detector with an impact parameter resolution of $5 \oplus 10/(p \beta \sin ^{3/2} \theta)~(\mu m)$.

 In addition to this impact parameter resolution the vertex detector needs to operate under a large amount of background. Most of the background events come from electron-positron events (pair background), which are generated at the IP due to beam-beam interactions. These pair background events cause hit occupancy in the vertex detector to increase, especially in the inner-most layer.
 
  A simulation study show that if we accumulate all the hits from one bunch train, the hit occupancy in the inner-most layer would be about $10\%$ for the pixel size of $25 \times 25~(\mu m^{2}) $.  
A hit occupancy around $1\%$ is needed for reasonable track reconstruction.

To satisfy this hit occupancy restriction two solutions are being considered. One is to readout multiple times, e.g. 20 times in one train.
The other is to use finely segmented pixel sensors as small as $5 \times 5~(\mu m ^{2} )$ and readout during the inter-train time, i.e. 200~ms ~\cite{ccd_glc, fpccd_option}. In this report, we describe a study of readout electronics for a realization of the latter option - Fine Pixel CCD(FPCCD). 

\section{Readout ASIC for FPCCD}
Developing a fully-depleted, radiation-hard $5 \times 5~(\mu m ^{2} )$ pixel sensor is challenging in itself and studies are currently being conducted. Furthermore, the implementation of FPCCD imposes
 several challenges for its readout electronics.  

 The first requirement is readout speed. In order to readout all  $128 \times 20,000$ pixels in one channel within the inter-train time, we need a readout speed over 10~Mpixels/s. Because we are using 10 bits to digitize one signal, this is equivalent to 10~ns/bit.
 
The second is noise level. As we reduce the pixel size the signal  level for a particle injected in a shallow angle decreases; therefore, to maintain signal integrity the noise level must be suppressed. The required noise level is 50 electron in total, among which 30 electrons are the requirement for the readout ASIC. 

The third requirement is power consumption. In order to keep the temperature of the vertex detector as low as $-40\hspace{-2pt} \ {}^\circ\mathrm{C}$, we must employ a cryostat. This restricts the power consumption of the vertex detector to be below 100~W. Since there are 6000 channels in the vertex detector this is equivalent to 16~mW/ch. The FPCCD sensor is estimated to consume approximately 10~mW/ch which means the readout ASIC is allowed to consume no larger than 6~mW/ch.

Readout precision is also required. Readout precision can be characterized with differential non-linearity and is required to be under a level compatible to electronic noise. 
Dynamic range up to 1500 electrons will be needed to accurately measure high level signals. 
\begin{table}[htbp]
\vspace{-1.5em}
\caption{summary of specification goals }
\label{tab:limits}
\centerline{\begin{tabular}{ | l  | l |  }
\hline
   &  Specification  goals    \\\hline \hline
Readout speed  & 10~ns/bit         \\\hline
Noise level  & 30 electrons     \\\hline
Power consumption  &   6~mW/ch        \\\hline
Dynamic range      &   1500 electrons \\\hline
Differential non-linearity	  & compatible to   electronic noise\\\hline
Ambient temperature	 &  $-40\hspace{-2pt} \ {}^\circ\mathrm{C}$ \\\hline
Noise shaping  & CDS \\\hline
On-chip digitization & charge sharing ADC \\
\hline
\end{tabular}}
\vspace{-0.7em}
\end{table}

\begin{wrapfigure}{t}{0.5\columnwidth}
\hspace{-1em}
\centerline{\includegraphics[width=0.45\columnwidth]{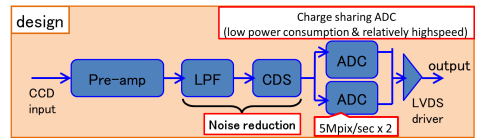}}
\caption{A block diagram of the circuit blocks in
each channel of the readout ASIC}\label{Fig:MV}
\end{wrapfigure}

\indent Our aim is to develop a readout ASIC which satisfies these requirements.
So far two prototypes have been produced. The design of these prototypes were focused on meeting the noise and readout speed requirements. 
 The specification goals are summarized in Table 1.  

The design we have adopted employs an amplifier, low pass filter (LPF), correlated double sampling (CDS), and  two charge sharing ADCs.
The block diagram for each channel is shown in Figure 1. 
Readout speed of 10~ns/bit is achieved by alternatively using two 20~ns/s charge sharing ADCs; where when one ADC is acquiring an analog signal, the other ADC is digitizing the signal and vice versa.  
Noise is reduced to 30 electrons by using LPF and CDS.
We have chosen a charge sharing successive approximation ADC to perform analog-to-digital conversion~\cite{adc}. The conversion process proceeds in terms of charge stored in the binary-weighted capacitors; which is quite different from the voltage-based A-to-D converter. It can operate under relatively high speed (20~ns/bit) and power consumption is low, which is ideal for the readout system for FPCCD.
\section{Problems seen in the First prototype}


\indent Two problems emerged after testing the first prototype. The first problem was that we weren't able to meet the readout speed requirement.
When operating under a readout speed exceeding 67~ns/bit, the distribution of the ADC output for a given test pulse input widened and several peaks were observed.
This is thought to be caused by a current shortage to the ADC comparator \cite{nima}.

The second problem was that large jumps were seen in the linearity curve of ADC count vs input voltage\cite{nima}. 
This can be explained through simulation to be the effect of stray capacitances attached to the switches and capacitors in the capacitor array. Since we have selected a charge sharing successive approximation ADC,
 each capacitance has to correspond to bit weight, thus the deviation between the ideal capacitance and the actual capacitance directly effects the ADC count. Large deviation leads to large jumps in the plot of ADC count vs input voltage.  
 
The measured noise level was 40 electrons when operating under the readout speed of 67~ns/bit at room temperature. Which is above the noise level requirement of 30 electrons.
\vspace{-0.6em}
\section{Design and results of the second prototype}


\indent In order to solve the problems seen in the first prototype, two major counter measures were taken in the second prototype.\\
\vspace{-1em}
\begin{figure}[htbp]

 \begin{minipage}{0.55\hsize}
  \begin{center}
\includegraphics[width=7cm]{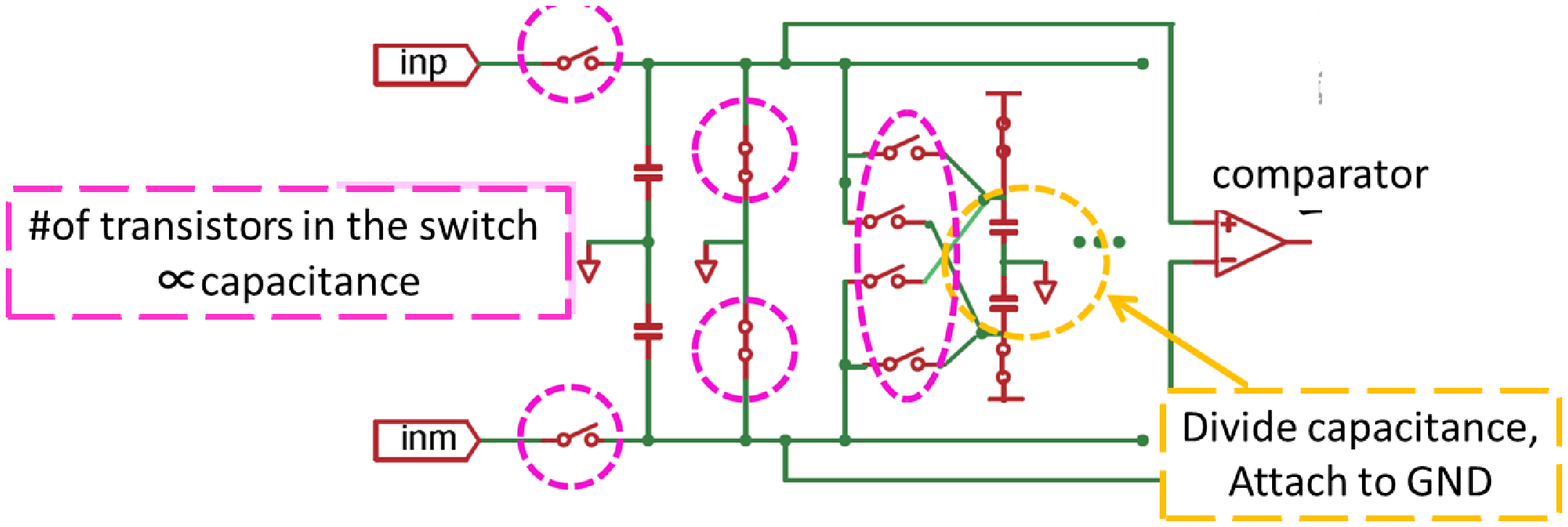}\label{fig:1st_adc}
\caption{circuit of charge sharing ADC}
\end{center} 
\end{minipage}
 \begin{minipage}{0.35\hsize}
 \begin{center}
 \includegraphics[width=5cm]{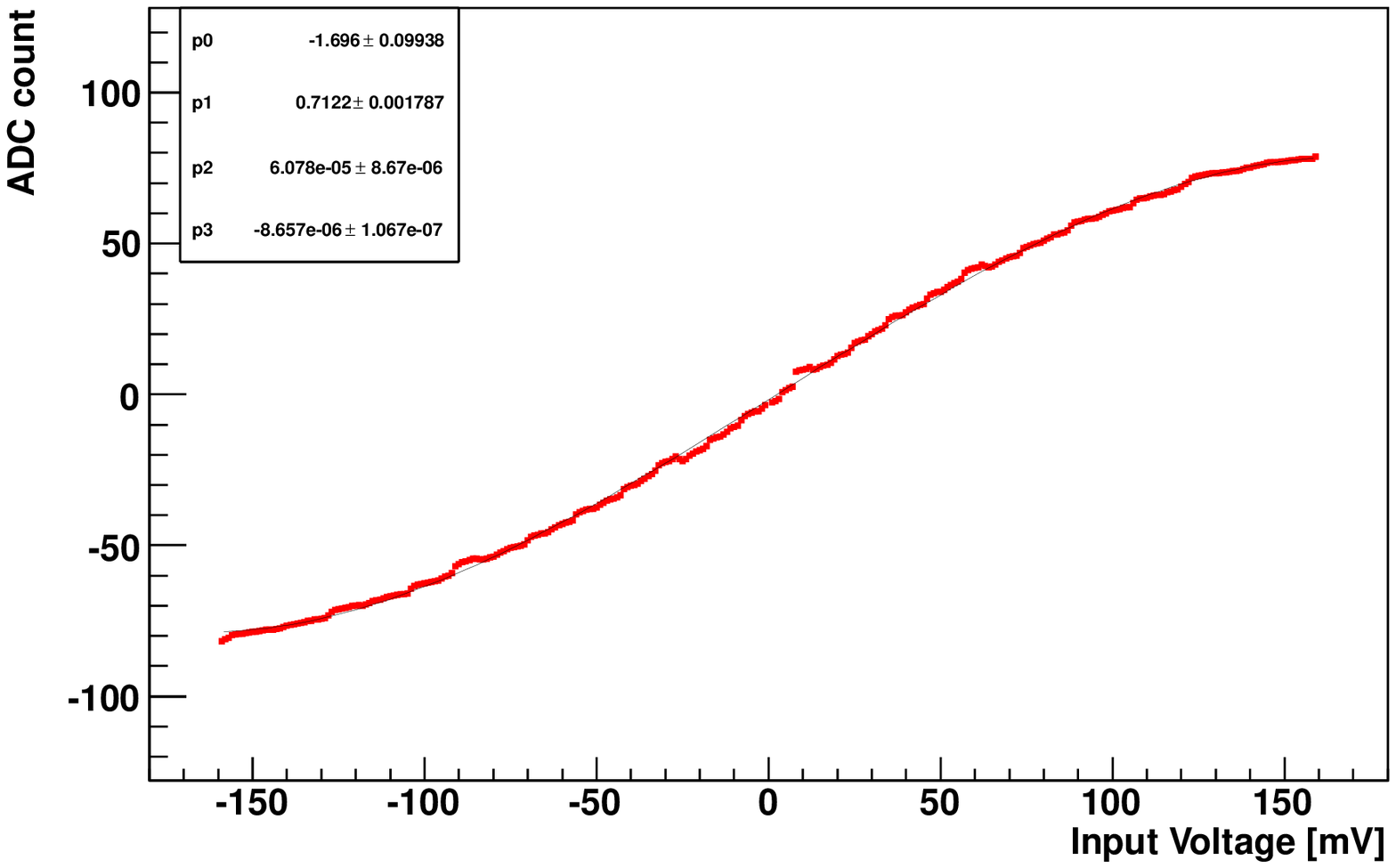}\label{fig:linearity}
 \caption{ADC count in relation to test pulse input voltage}
  \end{center}
  \end{minipage}
\end{figure}
\vspace{-1em}

To overcome the readout speed limitation problem, we increased the current supply to ADC comparator by increasing the numbers of pins from 80 to 100. 
For the problem of jumps in ADC count, we suppressed the effect of stray capacitance. 
As shown in the ADC circuit diagram in Figure 2, 
stray capacitance between capacitors and ground were suppressed by dividing each capacitor in the capacitor array into two and connecting the capacitor's bottom electrode to earth. 
Stray capacitance attached to the switches were suppressed by setting the number of transistors in the switches to be proportional to the binary weighted capacitors.  
Including these modifications we fabricated the second prototype with the TSMC 0.35~$\mu m $ mixed signal CMOS process. The prototype includes 8 readout channels on a 4.3 ~mm by 4.3~mm silicon area. 
First we verified performance of readout speed and succeeded in operating under 10~ns/bit; the required readout speed. 
Then we evaluated the degree of precision of the ADC output. Figure 3 shows relation between ADC count and input voltage of the test pulse.  
 There are no longer any large jumps in the ADC output which were seen in the first prototype. 

If we look further into Figure 3, a strong non-linear effect can be seen. This  non-linear effect can be quantified by integral non-linearity .

\subsection{Integral non-linearity}

\begin{table}[!hbt]\label{tab:inl}
	\caption[Integrated non-linearity and noise at different ranges]
	{\small Integrated non-linearity and noise at different ranges}
	\begin{tabular}{ | l  | l | l |}
	\hline
Input voltage range ~ [mV]  & Integral non-linearity$ ~[\%]$ &  Noise level  in CCD output ~$[\mu V]$ \\\hline \hline 
[-40:40] & 4.01$\%$& 45.5 \\\hline
[-80:80] &  6.58$\%$ &   48.5    \\\hline
[-160:160] &  17.08$\%$ &   52.5      \\
\hline
	\end{tabular}
\end{table}
Integral non-linearity(INL) is a measure which shows the global deviation from linearity. 
Although the allowance on integral non-linearity has not been specified so far, it is appropriate to reduce the amount of the integral non-linearity so as to utilize the full dynamic range of the A-to-D converter.  
Here we define INL as 
\begin{eqnarray}
\displaystyle INL \equiv | \frac{max(f(x)-g(x))}{g(x_{0})} | \times 100 \hspace{4pt} [\%]
\end{eqnarray}
where $x$ is the input voltage of the test pulse, the range of input voltage is [$-x_{0}$ :$x_{0}$], $f(x)$ is the fitted function. $g(x) $ is the linear function which runs through the origin and $f(x_{0})$  (or $f(-x_{0})$). 
Note that we are treating data after offset correction. 

The results of INL is shown in Table 2. 
The reason of the large integral non-linearity in our prototype was identified as due to saturation features from upstream circuits. 
Effects from upstream circuits can be seen even in narrow range. 
This can be improved in the next prototype. Detailed requirements to INL will be determined in the future tests of the detector.

\subsection{Noise}
Noise level can be extracted from the pedestal distribution. 
The extracted noise level which can be detected at the CCD output is shown in Table 2. Since the voltage per 1~LSB varies with input voltage range the calculated noise also changes. 
The numbers in the table show extracted noise level from pedestal distribution for three different ranges of input voltage. 
If we assume the gain in the CCD to be 5~$\mu V$ per electron, this will be equivalent to approximately 10 electrons. 
Which meets noise level required for the ASIC.

\subsection{Differential non-linearity}

 INL can be interpreted as a quantitative value used to evaluate the effects of upstream circuits prior to the ADC. 
To evaluate the precision of the ADC itself we would have to calculate differential non-linearity. 
Differential non-linearity(DNL) is the local deviation from the ideal function 
(in this case the fitted function f(x)).\\
\newpage
\begin{wrapfigure}{t}{0.5\columnwidth}

\centerline{\includegraphics[width=0.45\columnwidth ,height=4.5cm]{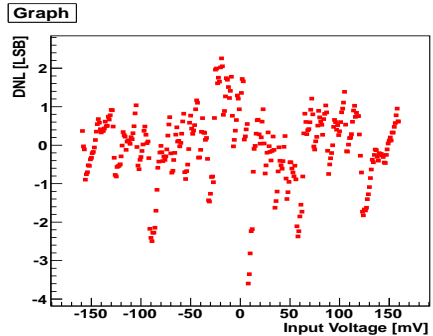}}
\caption{Differential non-linearity}\label{Fig:dnl}
\end{wrapfigure}
We define DNL as  
\begin{eqnarray}
 DNL \equiv f(x)-data(x)  ~[LSB]
\end{eqnarray}

\noindent where $data(x)$ is the ADC count for an input voltage $x$.
 The results of DNL are shown in Figure \ref{Fig:dnl}. The magnitude of DNL is approximately 6~LSB, which has the standard deviation of approximately 1.7~LSB. 
The noise level requirement is equivalent to 3.7~LSB. This indicates that the signal fluctuation is smaller than the required noise level.\\

\subsection{Power consumption}
The measured power consumption for the second prototype was 30.8~mW/ch. 
Which is significantly larger than the required power consumption of 6~mW/ch. 
The reason of the excessive power consumption was two sources: one is from redundant circuits which are provided to evaluate circuit performance, and the other is from the I/O buffers which draw static current irrelevant to the input logic level. 

\section{Summary and plan}
Based on the tests performed on the 1st prototype, we have made significant improvements for design of the second prototype; namely, we  
achieved the required readout speed of 10~ns/bit, we minimized any large jumps or missing bits in linearity and
reduced noise level to approximately 10 electrons which is well below the required amount.

Based on the performance study on the 2nd prototype, our aim towards the next prototype will be; 
to reduce power consumption, improve differential non-linearity and improve integral non-linearity.
Power consumption will be reduced by simplifying circuit design, cutting any DC current and reducing digital current by adopting 0.25~$\mu m$ process.
Differential non-linearity will be improved by stabilizing the ADC comparator through speed control. 
Although improvement of integral non-linearity is not essential upon satisfying the required specifications, dynamic range will be guaranteed by simplifying the CDS circuit. \\



\begin{footnotesize}


\end{footnotesize}


\end{document}